\documentclass[12pt]{iopart}
\usepackage{graphicx}

\begin{document}

\title[Direct measurement of the magnetic penetration depth in Ba(Fe$_{1-x}$Ni$_x$)$_2$As$_2$]{Direct measurement of the temperature dependence of the magnetic penetration depth in Ba(Fe$_{1-x}$Ni$_x$)$_2$As$_2$ superconductors}

\author{R I Rey$^1$, A Ramos-\'Alvarez$^1$, J Mosqueira$^1$, S~Salem-Sugui~Jr.$^2$, A D Alvarenga$^3$, H-Q Luo$^4$, X-Y Lu$^4$, R Zhang$^4$, F Vidal$^1$}

\address{$^1$LBTS, Facultade de F\'isica, Universidade de Santiago de Compostela, E-15782 Santiago de Compostela, Spain}

\address{$^2$Instituto de Fisica, Universidade Federal do Rio de Janeiro, 21941-972 Rio de Janeiro, RJ, Brazil}

\address{$^3$Instituto Nacional de Metrologia Qualidade e Tecnologia, 25250-020 Duque de Caxias, RJ, Brazil}

\address{$^4$Beijing National Laboratory for Condensed Matter Physics, Institute of Physics, Chinese Academy of Sciences, Beijing 100190, China}

\ead{j.mosqueira@usc.es}

\begin{abstract}
The temperature dependence of the in-plane magnetic penetration depth $\lambda_{ab}$ of Ba(Fe$_{1-x}$Ni$_x$)$_2$As$_2$ single crystals is determined directly from the \textit{shielding} magnetic susceptibility, measured in the Meissner region with the field parallel to the $ab$ layers. The doping levels studied cover the underdoped, optimally-doped and overdoped regimes. At temperatures below $0.5T_c$ a well-defined power-law behavior $\lambda_{ab}(T)-\lambda_{ab}(0)=AT^n$ (with $n\approx2.5$) is observed. At lower temperatures ($T<0.3T_c$) the data are still consistent with $n=2$ and $A\propto T_c^{-3}$, as predicted by the strong pair-breaking scenario proposed by Gordon \textit{et al}., Phys.~Rev.~B \textbf{81}, 180501(R) (2010). The temperature dependence of the superfluid density $\rho_s\propto\lambda_{ab}^{-2}$ presents a marked positive curvature just below $T_c$ which is a sign of two-gap superconductivity. The analysis of $\rho_s(T)$ in terms of a two-gap model allowed to estimate parameters like the in-band and inter-band couplings, the relative weight of each band, and their dependence with the doping level. A comparison with $\rho_s(T)$ data obtained by using other techniques in compounds with a similar composition is also presented.  
\end{abstract}

\submitto{\SUST}
\maketitle

\section{Introduction}

The study of the superconducting gap symmetry in Fe-based superconductors (FeSCs) may provide key information on the pairing interaction in these unconventional superconductors. So, works analyzing the differences in the gap structure among the FeSC families, and probing its evolution with the type and concentration of dopants, are at the forefront of the research in these materials.\cite{reviews} One of the two fundamental lengths in superconductors, the magnetic penetration depth $\lambda$, is directly related (through its temperature dependence) to the superconducting energy gap, and constitutes an useful tool to obtain information about its symmetry.\cite{Prozorov11}
The absolute value and temperature dependence of $\lambda$ has been recently investigated in FeSC by using different experimental techniques, including  
tunnel diode resonator (TDR),\cite{Fletcher09,Prozorov09,Tanatar09,Gordon09b,Gordon09a,Malone09,Martin09b,Martin09a,Kim10,Martin10b,Martin10a,Kim10b,Hashimoto10b,Gordon10a,Hashimoto10,Kim10a,Kim11,newCho11,Cho12b, Hashimoto12,Cho12a,Rodiere12,newMurphy13}
muon-spin rotation ($\mu$SR),\cite{Luetkens,Khasanov08,Williams09,Khasanov10,Bendele10,Shermadini10,newOfer12} microwave cavity perturbation,\cite{newOfer12}
two-coil mutual inductance,\cite{Yong11} the lower critical field ($H_{c1}$),\cite{Rodiere12,Choi10,newKlein10}
THz conductivity, \cite{Fischer10,Valdes10} surface impedance,\cite{Bobowski10,Hashimoto09,Hashimoto09b,newBarannik13} and local probes as magnetic force microscopy (MFM),\cite{Luan10,Luan11} scanning SQUID microscopy (SSS),\cite{Luan10,Luan11,Hicks09,newLippman12} and miniature Hall sensors.\cite{Rodiere12} Currently there is considerable consensus that the low-temperature behavior of $\lambda$ may be described by a power-law 
\begin{equation}
\lambda(T)=\lambda(0)+AT^n,
\label{powerlaw}
\end{equation}
where the exponent value depends on the particular FeSC family and, within the same family, on the type and concentration of the dopants. It is also well established that, in the most studied FeSC families (e.g., 111, 122, 1111), when phosphor is used as pnictide (completely or partially replacing arsenic) $n$ is close to 1, which is consistent with a nodal superconducting order parameter.\cite{Fletcher09,Hashimoto10b,Hashimoto12,newMurphy13,Hicks09} It has been suggested that this occurs when the pnictogen height from the iron plane decreases below $\sim1.33$~\r{A}.\cite{Hashimoto12} 

In the case that As is used as pnictide the situation is by far more complex. Exponents ranging from $n\approx1$ (in \mbox{NaFe$_{2-x}$Co$_x$As$_2$} away from optimal doping,\cite{Cho12a} and in clean \mbox{KFe$_2$As$_2$}\cite{Hashimoto10}) to $n\approx 3$ (in LiFeAs,\cite{Kim11} NdFe$_{1-x}$Co$_x$AsO,\cite{Kim10} and in optimally doped BaFe$_{2-x}$Co$_x$As$_2$\cite{Valdes10}) were reported. In some samples (e.g., LiFeAs,\cite{Hashimoto12} SmFeAsO$_{1-x}$F$_x$,\cite{Malone09} BaFe$_{2-x}$Co$_x$As$_2$,\cite{Luan11} and PrFeAsO$_{1-x}$\cite{Hashimoto09b}) it has been even found the exponential behavior typical of a fully-gapped superconducting order parameter (which may be parametrized by a power law with $n\stackrel{>}{_\sim}3$). Such a diversity has been attributed to differences from sample to sample of the pair-breaking scattering, which may change the clean-limit low-temperature behavior (power law with $n=1$, in the case of an order parameter with vertical line nodes, and exponential in the case that it is fully gapped) to a power law with $n\approx2$ in the dirty limit.\cite{Prozorov11,Gordon10b,Vorontsov09} However, to some extent the differences observed in the low temperature behavior of $\lambda(T)$ could also be attributed to the different experimental conditions of the techniques used. Some works report the agreement between different techniques when using the same sample: e.g., between $\mu$SR and microwaves in Ref.~\cite{newOfer12}, and between TDR and scanning SQUID in Ref.~\cite{newLippman12}. However, large differences observed between TDR and $H_{c1}$ measurements, and the dispersion between TDR measurements have been attributed to the sensitivity of this last technique to the edge roughness of the samples (see Ref.~\cite{newKlein10} and also Ref.~\cite{Hashimoto10}). This issue has recently been subject of discussion, see Ref.~\cite{newCho11} and the subsequent Comment and Reply (Ref.~\cite{commentreply}). As an example of the differences encountered when using different techniques, in optimally doped Ba(Fe$_{1-x}$Co$_x$)$_2$As$_2$ (one of the most widely studied compounds) $\lambda(T)$ follows Eq.~(\ref{powerlaw}) with $n=2-2.5$ by using TDR\cite{Prozorov11,Prozorov09,Gordon09b,Gordon09a} and 2-coil mutual inductance,\cite{Yong11} $n=2.8$ by using mm-wave surface impedance,\cite{newBarannik13} $n=3.1$ by using THz spectroscopy,\cite{Valdes10} and an exponential behavior is found by using local probes (MFM and SSS).\cite{Luan11} Large differences are also found in the full-range temperature dependence of the superfluid density $\rho_s\propto\lambda^{-2}$ in the same compound.\cite{Prozorov11,Gordon10a,Williams09,newOfer12,Yong11,Luan11,newBarannik13} 

Here we present new measurements of the temperature dependence of the in-plane magnetic penetration depth $\lambda_{ab}$ (the one associated to currents flowing in the crystal \textit{ab} layers),
\begin{equation}
\Delta\lambda_{ab}(T)\equiv\lambda_{ab}(T)-\lambda_{ab}(0),
\end{equation}
in a set of high-quality Ba(Fe$_{1-x}$Ni$_x$)$_2$As$_2$ single crystals with doping levels covering the underdoped, optimally-doped, and overdoped regimes.
These measurements were obtained from the \textit{shielding} magnetic susceptibility when a low external magnetic field (in the Meissner region) is applied parallel to the crystal \textit{ab} layers. To our knowledge this technique was not used before in FeSC,\footnote{This technique was previously used to determine the in-plane magnetic penetration depth in a high-$T_c$ cuprate superconductor, see Ref.~\cite{Krusin}. Some problems arising in the data analysis are commented on in Ref.~\cite{Krusincomment}.}
but unlike other techniques it allows the use of DC applied magnetic fields in the Oe range, and is very direct (the change in $\lambda_{ab}$ with the temperature is just proportional to the change in the measured magnetic moment). Also, it avoids well known difficulties associated with surface barriers (in the case of techniques based in the determination of $H_{c1}$)\cite{barriers} or the above mentioned problems associated with edge roughness (in the case of TDR). The typical size of the crystals used in the experiments lead to a diamagnetic moment in the $10^{-6}$~emu range, still well above the resolution of current SQUID magnetometers (see below). Martin \textit{et al}.\cite{Martin10b,Martin10a} and Rodi\`ere \textit{et al}.\cite{Rodiere12} already measured the $\Delta\lambda_{ab}$ behavior at low-temperatures in Ba(Fe$_{1-x}$Ni$_x$)$_2$As$_2$ with different doping levels by using the TDR technique. Both works report a power-law behavior at low temperatures, but find significant differences in the exponent, mainly in the overdoped region. Our measurements were obtained in the range $(0.1-1)T_c$, allowing to investigate the power-law behavior of $\Delta\lambda_{ab}$ at low temperatures ($\stackrel{<}{_\sim}0.3T_c$), but also the superfluid density at higher temperatures, still unexplored in the studied compound.

The experimental details are presented in Section 2. The low-temperature behavior of $\Delta\lambda_{ab}$ is analyzed in Section 3.1. The superfluid density, obtained by using the $\lambda_{ab}(0)$ values in the literature, is analyzed in the full temperature range below $T_c$ in Section 3.2. Finally, the concluding remarks are presented in Section 4.

\section{Experimental details and results}

The Ba(Fe$_{1-x}$Ni$_x$)$_2$As$_2$ crystals were grown by the self-flux method. Their nominal Ni doping levels are $x = 0.0375$ (underdoped, ud), 0.05 (optimally doped, op), and 0.075 (overdoped, ov). The details of the growth procedure and a thorough characterization may be found in Ref.~\cite{growth}. To avoid the complications associated with demagnetizing effects (see below) we used plate-like single crystals (typically $1\times1\times0.02$ mm$^3$, see Table I) with the FeAs ($ab$) layers parallel to their largest faces. They were cleaved from larger crystals by using adhesive tape. The surface irregularities are of the order of 50~nm in depth (as determined by AFM), and the uncertainty in the crystals thickness $L_c\approx20~\mu$m (the length relevant for the analysis) is below 1\%.

\begin{figure}[b]
\begin{center}
\includegraphics[scale=.7]{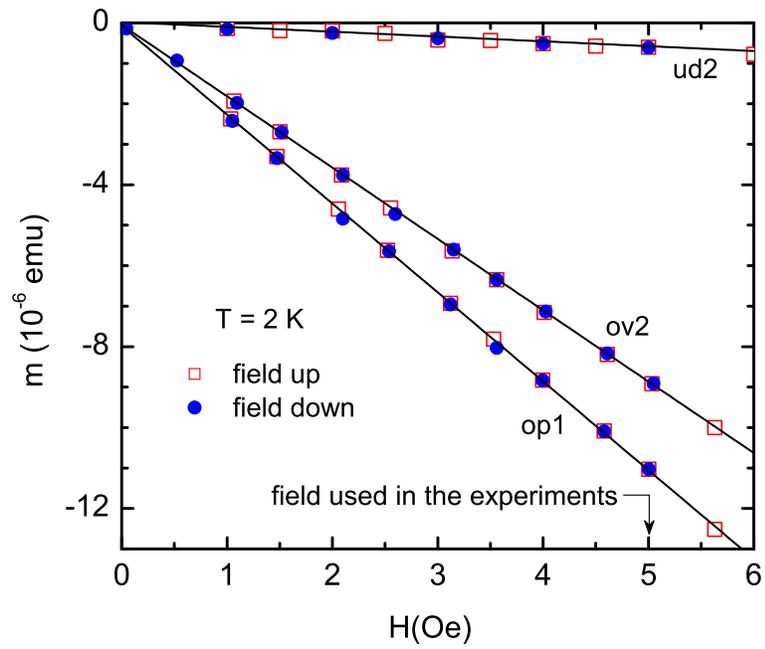}
\caption{Examples of the magnetic field dependence of the magnetic moment for all compositions studied. These measurements were performed with $H\parallel ab$ layers (i.e., parallel to the crystals largest faces) after a precise zero-field cooling to 2 K, see main text for details. The well defined linear and reversible behavior rules out the possible presence of spurious effects associated with magnetic flux trapped inside the crystals.}
\label{reversibility}
\end{center}
\end{figure}

The magnetic susceptibility with $H\parallel ab$, $\chi_\parallel$, was measured in several crystals of each composition with a Quantum Design's SQUID magnetometer (model MPMS-XL). For that we used a quartz sample holder (0.3 cm in diameter, 22 cm in length) to which the crystals were glued with a minute amount of GE varnish. Two plastic rods at the holder ends ($\sim0.3$ mm smaller than the sample space diameter) ensured an alignment better than $0.1^\circ$. However, the presence of the Ge varnish may introduce an additional uncertainty in the crystal orientation which effect will be commented below. The samples were zero-field cooled (ZFC) by using the \textit{ultra-low-field} option, which includes a shield for the earth's magnetic field and a conventional coil to compensate the superconducting coil's remnant field down to the $10^{-2}$~Oe level. The magnetic moment $m$ was measured against temperature (from 2~K up to above $T_c$) in presence of a 5~Oe applied magnetic field, which ensured a response linear and reversible. This is illustrated in Fig.~\ref{reversibility} where we present examples for all compositions studied of the $m(H)$ dependence upon increasing the field above 5 Oe, and then decreasing to zero. The well defined linear and reversible behavior rules out any spurious effects associated with magnetic flux trapping. Due to the small signal of the crystals studied (of the order of 10$^{-6}$~emu at low temperatures), we used the \textit{reciprocating sample option} (RSO) which performs sinusoidal oscillations of the sample about the center of the detection system and improves the sensitivity with respect to the conventional DC option. At each temperature we averaged 8 measurements consisting of 10 cycles at 1~Hz, the resulting uncertainty in magnetic moment being $\sim5\times10^{-9}$~emu. 

\begin{figure}[t]
\begin{center}
\includegraphics[scale=.7]{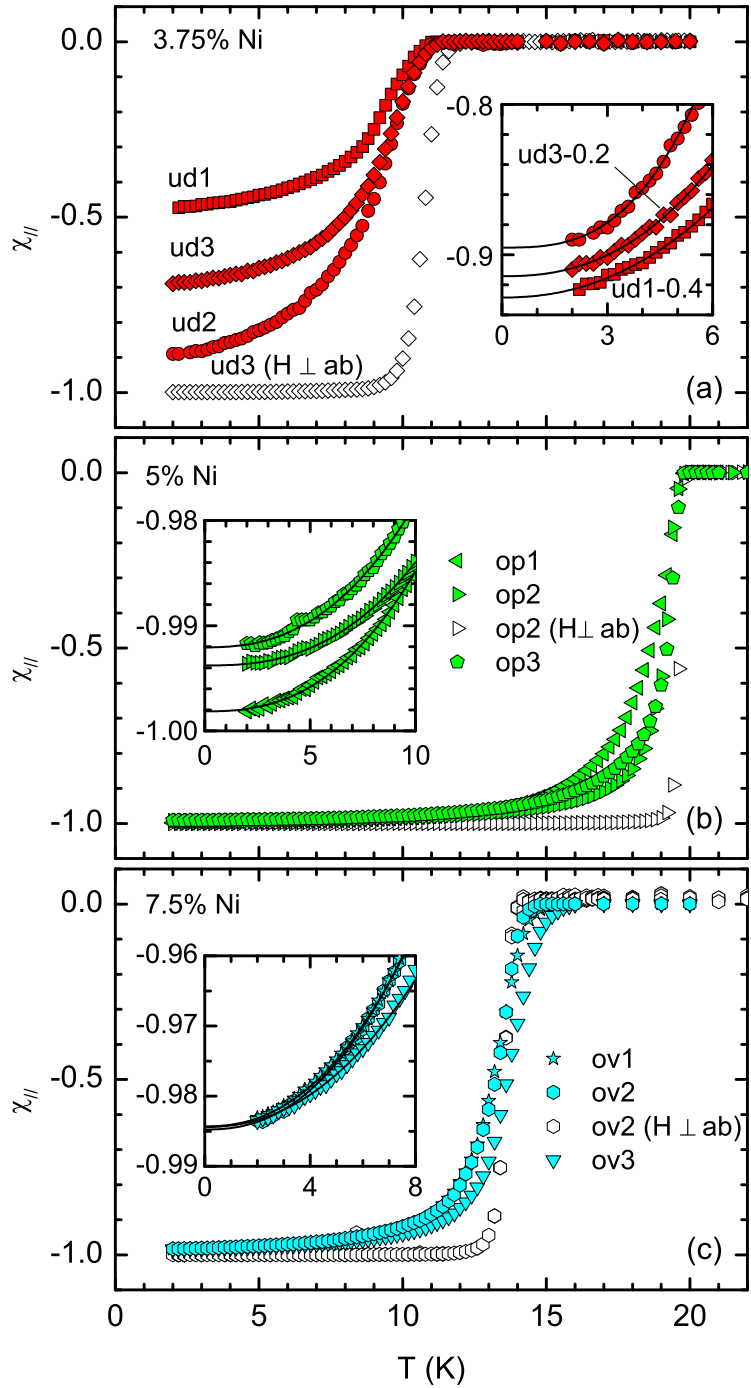}
\caption{Temperature dependence of the magnetic susceptibility of all samples studied, measured with $H=5$~Oe applied parallel to the Fe ($ab$) layers. White symbols were obtained with $H\perp ab$. Insets: Detail of the behavior at low temperatures (for clarity, the curves for crystals ud1 and ud3 with 3.75\% Ni are vertically displaced). The lines are fits of a power law for $T<0.5T_c$ to determine the low-temperature saturation value.}
\label{rawdata}
\end{center}
\end{figure}

The temperature dependence of $\chi_\parallel$ is presented in Fig.~\ref{rawdata}. From these curves, $T_c$ was estimated by linearly extrapolating to zero the higher-slope data, and the transition width as $\Delta T_c=T_{\rm onset}-T_c$, where $T_{\rm onset}$ is the highest temperature at which a diamagnetic signal is resolved. The resulting $\Delta T_c/T_c$ values (see Table~I) are among the best in the literature for crystals with the same composition.\cite{Prozorov11,Martin10b,Martin10a,Kim10a,Rodiere12,Gordon10b} For comparison, we include in Fig.~\ref{rawdata} some examples of the magnetic susceptibility obtained with $H\perp ab$, $\chi_\perp$ (open symbols). These last data were corrected for demagnetizing effects by using the demagnetizing factors $D_{\perp}$ needed to attain the expected value of $\chi_\perp=-1$ at low temperatures, which resulted to be consistent with the values that may be obtained from the crystals' dimensions\cite{Osborn} (the differences are within 5\%). While $\chi_\perp$ is temperature independent up to very close to $T_c$ (confirming the excellent quality of the crystals), $\chi_\parallel$ is notoriously rounded just below $T_c$ due to the competition of $\lambda_{ab}$ with the crystal thickness, $L_c$, on approaching $T_c$. 

\section{Data analysis}

\subsection{Low-temperature behavior of the in-plane penetration depth}

In view of the crystals geometry, the relationship between $\chi_\parallel$ and $\lambda_{ab}$ may be approximated by,\cite{Schoenberg}
\begin{equation}
\chi_\parallel=-1+\frac{2\lambda_{ab}}{L_c}\tanh\frac{L_c}{2\lambda_{ab}}.
\label{scho}
\end{equation}
This expression would allow to determine the absolute value of $\lambda_{ab}$ directly from the $\chi_\parallel$ data in Fig.~\ref{rawdata}. However, even a small crystal misalignment may lead to a non-negligible contribution coming from the field component perpendicular to the Fe-layers. Denoting $\alpha$ to the possible angle between $H$ and the crystal \textit{ab} layers, the measured magnetic susceptibility would be
\begin{equation}
\chi_\parallel^{\rm meas}=\frac{\chi_\parallel}{1+\chi_\parallel D_\parallel}\cos^2\alpha+\frac{\chi_\perp}{1+\chi_\perp D_\perp}\sin^2\alpha,
\end{equation}
where $D_\parallel$ is the demagnetizing factor for $H\parallel ab$. From the crystals dimensions in Table~1 it may be approximated $D_\parallel\approx \pi L_c/4L_{ab}\approx0.015$ and $D_\perp\approx1-2D_\parallel\approx0.97$.
% is found $D_\perp=0.965\pm0.010$ and $D_\parallel\approx(1-D_\perp)/2\approx0.017\pm0.005$. 
As $\chi_\perp\approx-1$ up to very close to $T_c$, this expression may be approximated by
\begin{equation}
\chi_\parallel^{\rm meas}\approx\chi_\parallel-\frac{\sin^2\alpha}{1-D_\perp},
\label{const}
\end{equation}
i.e., the difference between the measured magnetic susceptibility and the actual $\chi_{\parallel}$ is a temperature independent value (of the order of $10^{-2}$ for $\alpha\approx1^\circ$). While this may difficult determining the absolute value of $\lambda_{ab}$, it allows to determine its temperature dependence $\Delta\lambda_{ab}(T)=\lambda_{ab}(T)-\lambda_{ab}(0)$ with accuracy. As the crystals thicknesses $L_{c}$ are of the order of 20~$\mu$m, and the reported values of $\lambda_{ab}(0)$ are smaller than 1~$\mu$m (see below), it may be safely approximated $\tanh(L_c/2\lambda_{ab})\approx1$ up to very close to $T_c$ [typically for $T<0.9T_c$ it is found $\lambda_{ab}(T)<0.2L_c$ and the approximation is accurate within 1\%]. Then, from Eqs.~(\ref{scho}) and (\ref{const}) it follows
\begin{equation}
\Delta\lambda_{ab}(T)\approx\frac{L_{c}}{2}\left[\chi_\parallel^{\rm meas}(T)-\chi_\parallel^{\rm meas}(0)\right].
\label{deltalambda}
\end{equation}
Taking into account the above mentioned resolution of our measurement system and the geometry of the crystals used, our technique allows to detect changes in $\lambda_{ab}$ of the order of $\sim5$~nm, slightly larger than the typical resolution of the TDR technique (about 1~nm, see e.g., the figures in Ref.~\cite{Martin10b} with TDR measurements in the same compounds).

A detail of the $\chi^{\rm meas}_\parallel(T)$ behavior at low temperatures is presented in the insets of Fig.~\ref{rawdata}. The low-temperature saturation values were determined by fitting a power law in a temperature region up to $\sim 0.5T_c$ (solid lines). The resulting $\chi_\parallel^{\rm meas}(0)$ values are compiled in Table I. Optimally doped and overdoped crystals present $|\chi_\parallel^{\rm meas}(0)|$ values typically 0.02 larger that the ones expected in view of the $\lambda_{ab}(0)$ values in the literature ($260\pm50$~nm in crystals with 5\% Ni, and $340\pm60$~nm in crytals with 7.5\% Ni).\cite{Rodiere12} This is consistent with crystal misalignments of about $\alpha\sim1.5^\circ$. Crystals with 3.75\% Ni present a larger scattering in the $\chi_\parallel^{\rm meas}(0)$ values that may hardly be attributed to crystal misalignments: as for this composition $\lambda_{ab}(0)=450\pm80$~nm,\cite{Rodiere12} one would have to assume $\alpha\approx2^\circ-7^\circ$. The scattering in the $\chi_\parallel^{\rm meas}(0)$ values of UD crystals could then be attributed to a possible presence of non-superconducting domains in the crystals, maybe associated to the proximity of this doping level to the non superconducting phase.

\begin{figure}[h]
\begin{center}
\includegraphics[scale=.7]{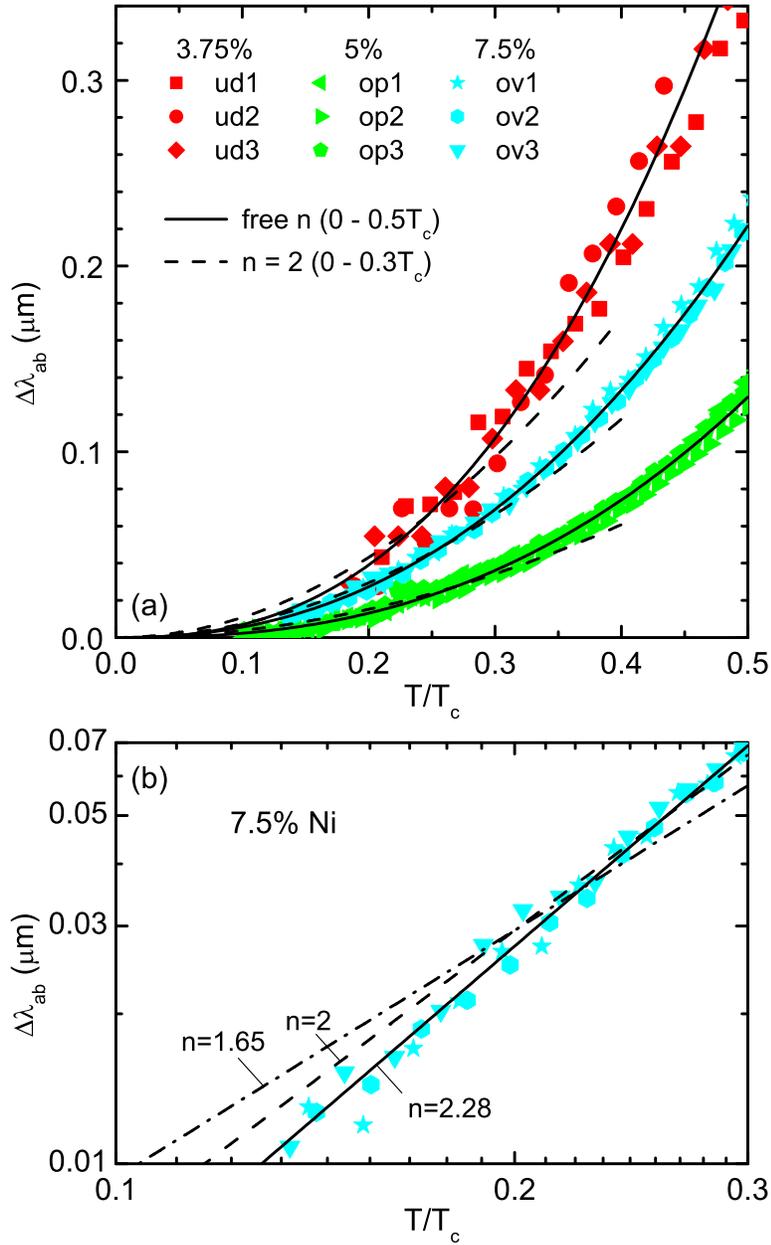}
\caption{(a) Low-temperature behavior of the in-plane magnetic penetration depth (in excess of the $T\to0$~K value) for all crystals studied. Solid lines are fits to a general power law up to $T/T_c=0.5$. Dashed lines are fits to a quadratic power law up to $T/T_c=0.3$. (b) Low-temperature detail in log-log scale for the overdoped crystals. Solid and dashed lines are the same as in (a). The dot-dashed line is a fit to a power law up to 0.3$T_c$ by using the $n$ value found in Refs.~\cite{Martin10a,Martin10b}.}
\label{lowT}
\end{center}
\end{figure}

\begin{figure}[t]
\begin{center}
\includegraphics[scale=.7]{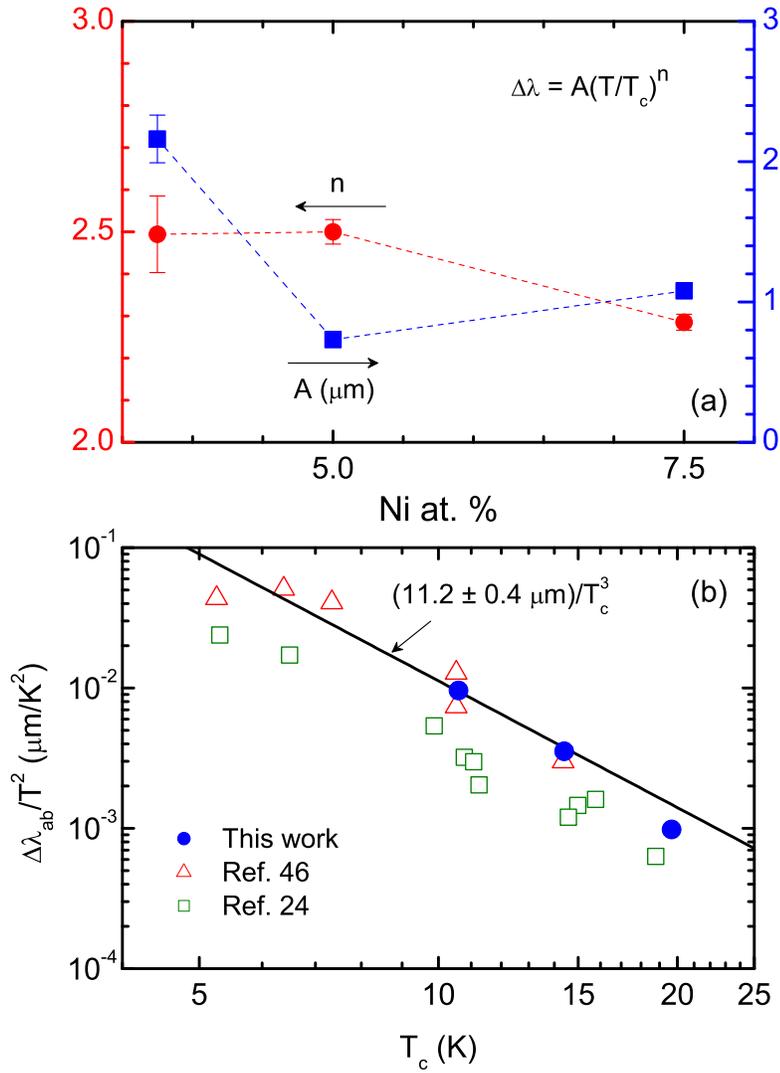}
\caption{(a) Amplitudes and exponents resulting from the fit of a general power law to the data in Fig.~\ref{lowT}. (b) Amplitude resulting from the fit of a quadratic power law to the data in Fig.~\ref{lowT} for $T<0.3T_c$. The solid line is a fit to the $T_c^{-3}$ dependence predicted by the strong pair-breaking approach developed in Ref.~\cite{Gordon10b}.}
\label{parameters}
\end{center}
\end{figure}

The low-temperature behavior of $\Delta\lambda_{ab}(T)$ for all crystals studied is presented in Fig.~\ref{lowT}. The data corresponding to crystals with the same composition roughly fall on the same curve, even in the case of the crystals with 3.75\% Ni.\footnote{The consistency between the $\Delta\lambda_{ab}(T)$ data in the three underdoped crystals justifies the applicability of Eq.~(\ref{scho}) also in these samples in spite that, as commented above, they may present a distribution of non superconducting domains. This may be explained by taking into account that non-superconducting domains in the samples interior are completely screened and have little effect in the measured ZFC magnetic susceptibility. The global $|\chi_\parallel|$ reduction observed in some of these samples may then be attributed to the presence of interconnected non-superconducting domains leading to large (of the order of the samples thickness) unscreened areas within the sample. Provided that they are much larger than $\lambda_{ab}$, its presence would not appreciably affect the $\chi_\parallel$ temperature dependence given by Eq.~(\ref{scho}).} The solid lines are fits of a power law, 
\begin{equation}
\Delta\lambda_{ab}(T)=A\left(\frac{T}{T_c}\right)^n,
\end{equation}
to the set of data for each composition up to $T/T_c=0.5$. The fit qualities are excellent in all the reduced-temperature range, and lead to the amplitudes and exponents shown in Fig.~\ref{parameters}(a). $n$ is about $\sim2.5$ up to the optimal-doping level, and decreases to $\sim2.3$ for $x=0.075$.\footnote{Note that for the underdoped samples the temperature range is restricted to $T>0.2T_c$, and the conclusions about the low-temperature behavior are less robust than in optimally-doped and overdoped samples.} It has been proposed that impurity scattering would strongly affect the low-temperature behavior of $\Delta\lambda_{ab}$. In particular, superconductors with a fully-gapped order parameter or with a $d$-wave symmetry, would change their exponential/linear temperature dependences in the clean limit to a power law with $n$ approaching 2 in the dirty limit.\cite{Prozorov11} Our present results, with $n$ values slightly above $n=2$, would be then consistent with a nodeless order parameter for all doping levels, affected by the presence of impurity scattering.  

It has been calculated that in the dirty limit the amplitude in the quadratic power law should be proportional to $T_c^{-3}$.\cite{Gordon10b} A fit to the experimental data by fixing $n=2$ (dashed lines in Fig.~\ref{lowT}(a)) is still reasonably good up to $T/T_c\approx0.3$. Also, as it is shown in Fig.~\ref{parameters}(b), the resulting amplitude follows the the predicted $T_c^{-3}$ dependence, and is close to the values found in several FeSC families including Ba(Fe$_{1-x}$Ni$_x$)$_2$As$_2$.\cite{Rodiere12,Gordon10b} This reinforces our above conclusion, and suggests that our crystals are close to the limit of strong impurity scattering. 

Our results are consistent with recent TDR measurements in crystals with similar compositions.\cite{Rodiere12} They are also coherent with the results of Kim \textit{et al.}\cite{Kim10a}, which showed that in optimally-doped Ba(Fe$_{1-x}$Ni$_x$)$_2$As$_2$ $n$ diminishes from $n=2.5$ to $n\sim2$ when defects are progressively introduced in the crystals through heavy-ion irradiation. However, there is a notable difference with the TDR measurements by Martin \textit{et al.}\cite{Martin10b,Martin10a} who find that $n$ falls significantly below 2 in similarly overdoped crystals from the 122 family (BaFe$_{2-x}$M$_x$As$_2$ with M=Pd, Co, Co+Cu, and also Ni). This result led these authors to suggest that the superconducting gap is not universal even within the same 122 family, and that in the overdoped regime it may become highly anisotropic and nodal. In agreement with this proposal, measurements of the fluctuation-induced magnetoresistance above $T_c$ in crystals from the same batches as the ones used in the present work,\cite{Deltasigma} showed that the superconducting anisotropy factor increases with $x$ from $\gamma\approx2$ at optimal-doping up to $\gamma\approx10$ at high doping levels (7.5\% Ni). In addition, studies of the low-temperature specific heat in in Ba(Fe$_{1-x}$Co$_x$)$_2$As$_2$,\cite{overnodes1} and of point-contact Andreev reflection in Ba(Fe$_{1-x}$Ni$_x$)$_2$As$_2$,\cite{overnodes2} also suggest the possible presence of nodes in the overdoped region. However, as it is shown in the detailed log-log representation of Fig.~\ref{lowT}(b), exponents below $n=2$ are clearly out of the experimental uncertainty in all our overdoped crystals.

\subsection{Temperature dependence of the superfluid density}

\begin{figure}[t]
\begin{center}
\includegraphics[scale=.7]{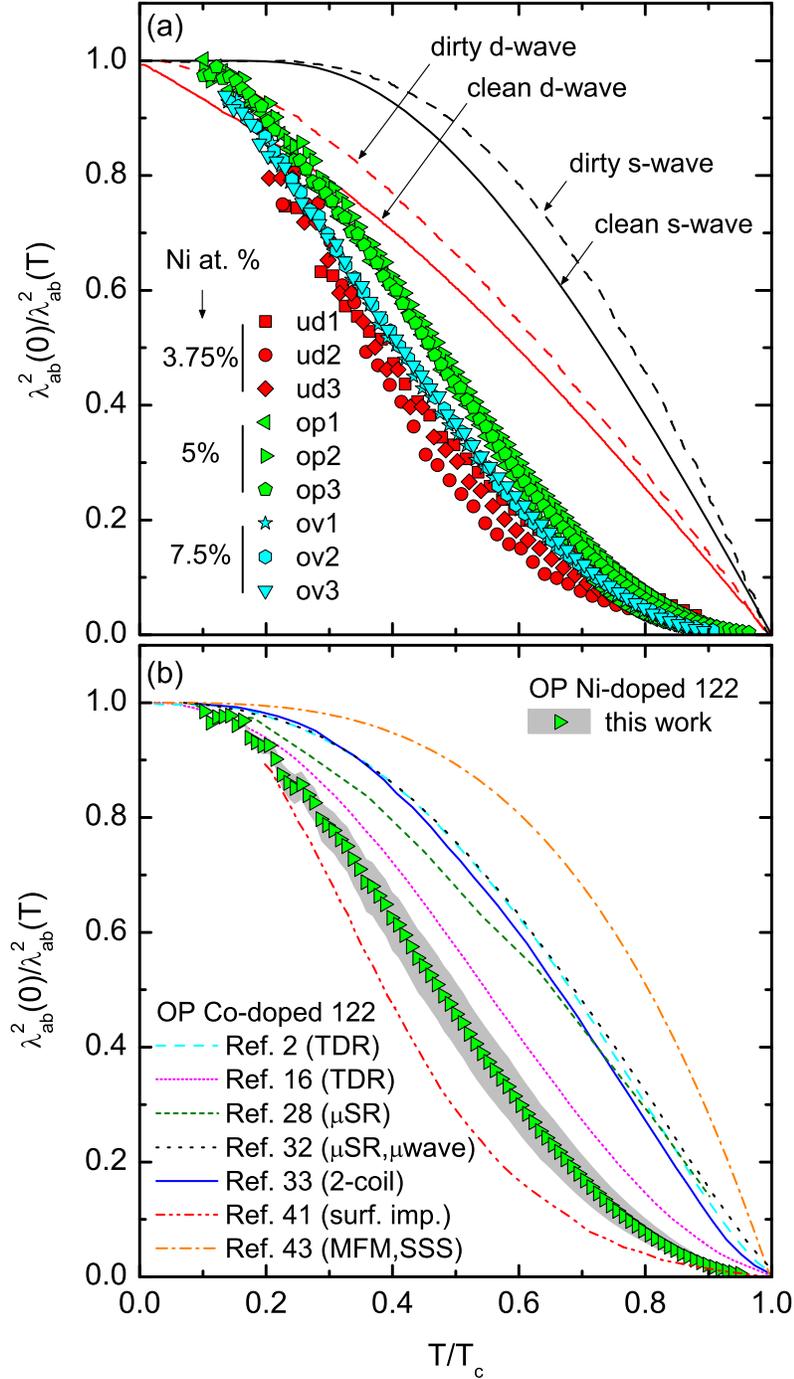}
\caption{(a) Reduced-temperature dependence of the normalized superfluid density as obtained from Eq.~(\ref{superfluid}) by using the $\Delta\lambda_{ab}(T)$ data resulting from Eq.~(\ref{deltalambda}) and the $\lambda_{ab}(0)$ data in Ref.~\cite{Rodiere12}. We only include data verifying $\lambda_{ab}(T)<0.2L_c$, for which Eq.~(\ref{deltalambda}) is correct within 1\%. For comparison, the results for single band $s$-wave and $d$-wave superconductors are also included. (b) Comparison between the normalized $\rho_s(T)$ data for optimally-doped Ba(Fe$_{1-x}$Ni$_x$)$_2$As$_2$ (corresponding to the crystal op2), and for optimally-doped Ba(Fe$_{1-x}$Co$_x$)$_2$As$_2$ (taken from the indicated works). The shaded area stands for the uncertainty in the $\lambda_{ab}(0)$ value used.}
\label{figsuperfluid}
\end{center}
\end{figure}

A more complete analysis of the superconducting gap structure may be done through the temperature dependence of the normalized superfluid density in the complete temperature range below $T_c$. It may be obtained through,
\begin{equation}
\frac{\rho_s(T)}{\rho_s(0)}=\left(1+\frac{\Delta\lambda_{ab}(T)}{\lambda_{ab}(0)}\right)^{-2}.
\label{superfluid}
\end{equation}
In this expression we used the $\lambda_{ab}(0)$ vs $T_c$ dependence for the Ba(Fe$_{1-x}$Ni$_x$)$_2$As$_2$ system derived in Ref.~\cite{Rodiere12} from local Hall magnetometry and, independently, from specific heat measurements. It leads to $\lambda_{ab}(0)=450\pm80$~nm for 3.75\%~Ni, $\lambda_{ab}(0)=260\pm50$~nm for 5\%~Ni, and $\lambda_{ab}(0)=340\pm60$~nm for 7.5\%~Ni. These values are consistent with the ones obtained in Ref.~\cite{Wu10} from optical reflectometry, and with the ones for Ba(Fe$_{1-x}$Co$_x$)$_2$As$_2$ with equivalent electron concentrations obtained in Ref.~\cite{Gordon10a} from TDR. The resulting $\rho_s(T)/\rho_s(0)$ is presented in Fig.~\ref{figsuperfluid} where, for comparison, it is also included the result for single-band $s$-wave and $d$-wave superconductors. Contrary to these conventional scenarios, the superfluid density presents a notable positive curvature for temperatures just below $T_c$ for all studied doping levels, which is a sign of two-gap superconductivity.\cite{Kogan09} This curvature is more pronounced in the overdoped and underdoped crystals, although the differences are close to the uncertainties in the $\lambda_{ab}(0)$ values (a representative example for one of the optimally-doped crystals is shown as a shaded area in Fig.~\ref{figsuperfluid}b). 

\begin{figure}[t]
\begin{center}
\includegraphics[scale=.7]{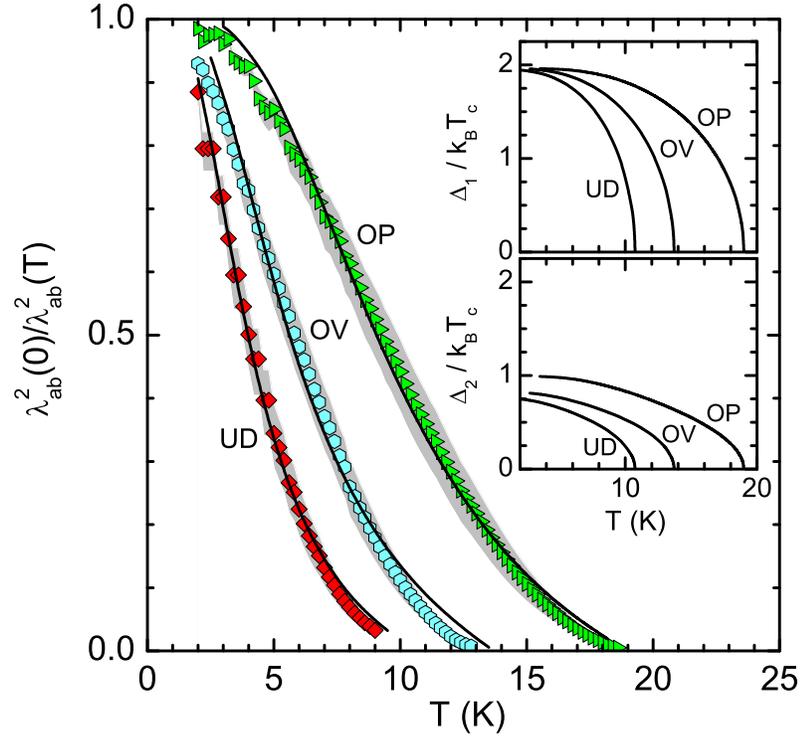}
\caption{Comparison of the superfluid density in our Ba(Fe$_{1-x}$Ni$_x$)$_2$As$_2$ samples (the data are examples corresponding to crystals ud3, op2 and ov2) with the two-gap gamma model of Kogan \textit{et al}. \cite{Kogan09} (solid lines). We only include data verifying $\lambda_{ab}(T)<0.2L_c$, for which Eq.~(\ref{deltalambda}) is correct within 1\%. The shaded areas stand for the uncertainties in the $\lambda_{ab}(0)$ values used. The resulting temperature-dependent superconducting gaps, $\Delta_1(T)$ and $\Delta_2(T)$, are presented in the insets. For details, see the main text.}
\label{theory}
\end{center}
\end{figure}

A quantitative analysis of our $\rho_s(T)/\rho_s(0)$ data is presented in Fig.~\ref{theory} in the framework of a self-consistent isotropic $s$-wave two-gap model (the so-called \textit{gamma model}).\cite{Kogan09} This model depends on parameters like the in-band ($\lambda_{11}$ and $\lambda_{22}$) and inter-band ($\lambda_{12}$) couplings, the relative density of states ($n_1$ and $n_2=1-n_1$), and the parameter $\gamma$ determining the partial contribution to the superfluid density from each band, $\rho_s=\gamma \rho_{s,1}+(1-\gamma)\rho_{s,2}$. This clean $s$-wave model should not work at low temperatures, where we observe a power-law behavior, but it is expected to provide a reasonable description at higher temperatures.\cite{Kim10b} To limit the number of the fitting parameters, we have considered that both bands have the same partial density of states, $n_1=n_2=0.5$, while $\lambda_{11}$ was set to give the correct $T_c$, assuming a Debye temperature of $150$~K.\cite{Kant} The lines in the main panel of Fig.~\ref{theory} are the fits to the $\rho_s(T)/\rho_s(0)$ data for temperatures above $\sim0.2T_c$. The resulting fitting parameters are presented in Table 2, and the corresponding temperature-dependent superconducting gaps, $\Delta_1(T)$ and $\Delta_2(T)$, are presented in the insets. As in other FeSC's $\Delta_1/\Delta_2\approx2$ and, in view of the small $\gamma$ value, the main contribution to $\rho_s$ comes from the band with a smaller gap.\cite{Kim10b,Kim11,Cho12b} However, the existence of the larger gap and a small interband coupling is needed to account for the high $T_c$.

To our knowledge, there are no $\rho_s(T)$ data for Ba(Fe$_{1-x}$Ni$_x$)$_2$As$_2$ to compare with. However, some works studied it in the very similar Ba(Fe$_{1-x}$Co$_x$)$_2$As$_2$. In the TDR measurements by Gordon \textit{et al}. in these compounds,\cite{Gordon10a} an analogous positive curvature was also observed near $T_c$ (see the pink dotted line in Fig.~\ref{figsuperfluid}b for the optimal doping level). In the same work it is also found a similar dependence of $\rho_s(T)/\rho_s(0)$ with the doping level, which was attributed to an enhanced gap anisotropy when departing from the optimal doping. However, authors using other techniques obtain results significantly different in the same compounds.\cite{Williams09,newOfer12,Yong11,newBarannik13,Luan11} As an example, in Fig.~\ref{figsuperfluid}b we compare the data in the literature for optimally doped Ba(Fe$_{1-x}$Co$_x$)$_2$As$_2$, with our data for Ba(Fe$_{1-x}$Ni$_x$)$_2$As$_2$. As it may be seen, in data from Refs.~\cite{Williams09}, \cite{newOfer12}, \cite{Yong11}, and \cite{Luan11} the positive curvature is much less pronounced or even is not observed, while in the most recent Ref.~\cite{newBarannik13} it is larger. It is also significant the notable differences observed by using the same technique in crystals coming from different batches (see Refs.~\cite{Prozorov11,Gordon10a}), which was attributed to differences in the impurity scattering between the samples used.

\section{Conclusions}

We have shown that measurements of the shielding magnetic susceptibility in the Meissner region allow a direct and reliable determination of the temperature dependence of the in-plane magnetic penetration depth in easily exfoliable Fe-based superconductors. By using high-quality Ba(Fe$_{1-x}$Ni$_x$)$_2$As$_2$ single crystals we studied the evolution of $\Delta\lambda_{ab}(T)\equiv\lambda_{ab}(T)-\lambda_{ab}(0)$ with the doping level. At low temperatures we find that it follows a power-law with an exponent $n\stackrel{>}{_\sim}2$ almost independent of the doping level, and an amplitude roughly proportional to $T_c^{-3}$. These results are consistent with a nodeless order parameter in the limit of strong impurity scattering. This is in agreement with results obtained by using other experimental procedures in the same compounds,\cite{Kim10a,Rodiere12} but contrasts with the results of Martin \textit{et al}.\cite{Martin10b,Martin10a} who find an exponent significantly below 2 in the strongly overdoped regime, consistent with an order parameter with line nodes in this region.   

When combined with the $\lambda_{ab}(0)$ values in the literature, our measurements also allowed to study the temperature dependence of the superfluid density $\rho_s\propto\lambda_{ab}^{-2}$ in the full temperature range below $T_c$. We find a marked positive curvature of $\rho_s$ in a wide temperature region below $T_c$ (slightly increasing for doping levels away from the optimal one) which is interpreted in the framework of a self-consistent isotropic $s$-wave two-gap model (the so-called \textit{gamma model}) \cite{Kogan09}. These results agree with the TDR measurements of Gordon \textit{et al}. \cite{Gordon10a} in the very similar Ba(Fe$_{1-x}$Co$_x$)$_2$As$_2$ system, but present notable differences with other works in the same compound.\cite{Prozorov11,Williams09,newOfer12,Yong11,newBarannik13,Luan11} More measurements are needed in order to determine the evolution of the gap symmetry with the doping level and across the different FeSC families.

\ack

Supported by the Spanish MICINN and ERDF \mbox{(No.~FIS2010-19807)}, and by the Xunta de Galicia (Nos.~2010/XA043 and 10TMT206012PR). SSS and ADA acknowledge support from CNPq and FAPERJ. The work at IOP, CAS in China is supported by NSFC Program (No. 11374011) and MOST of China (973 project: 2011CBA00110).

\newpage

%
% Table 1
%
\begin{table}[t]
\begin{center}
\begin{tabular}{cccccc}
\hline
Ni \%  & crystal & dimensions & $T_c$ & $\frac{\Delta T_c}{T_c}$ & $\chi^{\rm meas}_\parallel(0)$  \\
 & & (mm$^3$) & (K) &  &  \\
\hline
3.75  & ud1 & $0.75\times0.4\times0.015$ & 10.5 & 0.047  & -0.479 \\
 & ud2 & $0.5\times0.35\times0.010$ & 10.6 & 0.037   & -0.895 \\
 & ud3 & $0.9\times0.55\times0.013$ & 10.7 & 0.043   & -0.694 \\ \hline
5  & op1 & $1.7\times0.9\times0.021$ & 19.7 & 0.005 &  -0.998 \\
 & op2 & $1.35\times1.1\times0.027$ & 19.6 & 0.010  &  -0.994 \\
 & op3 & $1.2\times1.0\times0.020$ & 19.7 & 0.005  &  -0.992 \\
 \hline
7.5   & ov1 & $0.65\times1.15\times0.021$ & 14.3 & 0.076 & -0.985 \\
  & ov2 & $1.2\times1.5\times0.021$ & 14.1 & 0.049  & -0.984 \\
  & ov3 & $0.6\times1.15\times0.024$ & 14.8 & 0.080  & -0.984 \\\hline
\end{tabular}
\end{center} 
\caption{Some parameters of the crystals studied relevant for the analysis. See main text for details.}
\end{table}

%
% Table 2
%
\begin{table}[t]
\begin{center}
\begin{tabular}{cccccc}
\hline
Ni \%  & $n_1$ & $\lambda_{11}$ & $\lambda_{22}$ & $\lambda_{12}$ & $\gamma$  \\
\hline
3.75  & 0.5 & 0.68 & 0.32 & 0.120  & 0.02 \\
5  & 0.5 & 0.84 & 0.49 & 0.148 &  0.08 \\
7.5   & 0.5& 0.75 & 0.35 & 0.148 & 0.07 \\
\hline
\end{tabular}
\end{center} 
\caption{Parameters arising in the analysis of the superfluid density in the framework of the two-gap gamma model. See main text for details.}
\end{table}

%
% fig. 1
%

%
% fig. 2
%

%
% fig. 3
%

%
% fig. 4
%

%
% fig. 5
%

%
% fig. 6
%

\end{document}